\documentstyle[preprint,aps]{revtex}
\begin{document}
\pagestyle{plain}
\title{Mass Formula for a Stationary Axisymmetric Configuration
and the Physical Realization of the Kerr Metric}
\author{ R. M. Avakian $^a$ and G. Oganessyan $^{a,\
b,\ }$ \thanks{Corresponding author}}
\address{$^a$ Dept. of Theoretical Physics, Yerevan State
University, Yerevan 375049, Armenia}
\address{$^b$ Theoretical Astrophysics Group, Tata Institute of
Fundamental Research, Bombay-400 005, India \thanks{Address for correspondence.
E-mail: gurgen@tifrvax.tifr.res.in}}
\maketitle

\begin{abstract}

We analyse the expression for the mass of a stationary
axisymmetric configuration in general relativity obtained in our
previous work [1]. From the generality of our formula and its incompatibility
with the corresponding expression in Kerr space-time we argue
that a stationary equilibrium distribution of a real matter
cannot be a source of the Kerr metric.

\end{abstract}
\vskip 1cm

There exists an expression relating the mass of a stationary axisymmetric
configuration to the pressure distribution inside it [1]. The
line element for the metric created by such a configuration is written in the
coordinates which in the spherically
symmetric limit, for instance when the angular
velocity $\Omega$ becomes zero, go over into the isotropic form:
\newcommand{\begar}{\begin{eqnarray}}
\newcommand{\enar}{\end{eqnarray}}
\newcommand{\begeq}{\begin{equation}}
\newcommand{\eneq}{\end{equation}}
\begar
ds^2 =& ( e^\nu - \omega^2r^2\sin^2\theta e^\mu)\, dt^2 - e^\lambda
(dr^2 + r^2  d\theta^2 ) - r^2\sin^2\theta e^\mu d\phi^2 \\
\nonumber & -
2\omega r^2\sin^2\theta e^\mu  d\phi \, dt,
\enar
where $\nu,\mu,\lambda$ and $\omega$  are functions
of $r$, $\theta$ and $\Omega$.

  Since the components of the metric tensor do
not depend on $x^0 = c t$ and $x^3 = \phi$,  the components
 $R_0^0$ and $R_3^3$ of the Ricci tensor can be written as
\begeq
R_0^0 = {1 \over \sqrt{-g} } {\partial{} \over \partial{x^\alpha}}
	   (\sqrt{-g}\, g^{0 i} \Gamma_{0i}^\alpha),
\eneq
\begeq
R_3^3 = {1\over\sqrt{-g} } {\partial{}\over\partial {x^\alpha}}
	   (\sqrt{-g}\,g^{3 i} \Gamma_{3i}^\alpha),
\eneq
where $ g= det\, g_{i k} = - r^4 \sin^2\theta e^{\nu + 2\lambda +\mu},
 \,i= 0,1,2,3,\, \alpha =1,2,3$.
This covariant divergence form of the components of the Ricci tensor
allows us to apply Gauss's theorem and  use only the
asymptotic form of the metric functions while integrating expressions
containing $R_0^0$ and $R_3^3$ over the entire space.

{}From Einstein equations we then obtain the following mass
formula [1]:
\begeq
M^2 = - {32} \int_{0}^{\pi}\int_{0}^{{r_s}(\theta)}
( T_1^1 + T_2^2 ) \,r^3 \sin^2\theta e^{{\nu + 2\lambda + \mu}\over2}
 dr d\theta,
\eneq
where $T^1_1$ and $T^2_2$ are the corresponding components of
the energy-momentum tensor,
 and ${r_s}(\theta)$ is the boundary of the configuration,
 $T_1^1[{r_s}(\theta)] = T_2^2[{r_s}(\theta)] = 0 .$

In the case of rotating ideal fluid $(4)$ reduces to
\begeq
M^2 =  {64} \int_{0}^{\pi}\int_{0}^{{r_s}(\theta)}
P \,r^3 \sin^2\theta e^{{\nu + 2\lambda + \mu}\over2}
 dr d\theta.
\eneq
The derivation of $(4)$ is very general, the only assumption being
that the configuration is stationary ( equations $(2)$ and $(3)$
hold) and equlibrium ($T_1^1$ and $T_2^2$ are nonzero). One could try to
proceed in
the same manner and derive a similar formula for the Kerr metric,
which  also can be written in isotropic coordinates [5]:
\begeq
ds^2 = -X^2 dt^2 + r^2 \sin^2 \theta \frac{B^2}{X^2} (d\phi -
\omega dt)^2 + \frac{\psi^2}{X^2} (dr^2 + r^2 d \theta^2)
\eneq
The metric functions in $(5)$ are given by
\begeq
X^2 = \frac{\Sigma \Delta}{A}
\eneq
\begeq
\omega = 2 a \frac{R}{A} M
\eneq
\begeq
\psi^2 = \frac{\Sigma^2 \Delta}{r^2 A}
\eneq
\begeq
B^2 = \frac{\Delta}{r^2}
\eneq
where
\begeq
R = \frac{(2r + M)^2 - a^2}{4r}
\eneq
\begeq
\Sigma = R^2 + a^2 \cos^2 \theta
\eneq
\begeq
\Delta = R^2 - 2 R M + a^2
\eneq
\begeq
A = (R^2 + a^2)^2 - a^2 \Delta \sin^2 \theta,
\eneq
$M$ is the mass and $a = L/M$ is the specific angular momentum
of the black hole.

 Now, following the procedure by which $(4)$ was derived, we
obtain the following mass formula:
\begeq
M^2 - a^2 = - {32} \int_{0}^{\pi}\int_{0}^{{r_s}(\theta)}
 (T^1_1 + T^2_2) \,r^3 \sin^2\theta e^{{\nu + 2\lambda + \mu}\over2}
 dr d\theta,
\eneq
 Formula $(15)$ as  compared to $(4)$ has an extra term $-a^2$.
This difference is essential and cannot be neglected, except for
the trivial case of $\Omega = 0$, when both the line elements
(1) and (6) reduce to the Schwarzschild form. This suggests that
the conditions used in the derivation of $(4)$, namely that the configuration
of a real matter must be equilibrium and stationary, do not hold in
the case of the Kerr metric.

In a less general form this result has been known from the work
of one of the authors [$2$], where it is shown that, in the case
of a real gas of baryons, no coordinate transformation can bring
the exterior solution for a stationary axisymmetric stellar
configuration  in $\Omega^2$ approximation, known as
Hartle-Thorne-Sedrakian-Chubarian solution [$3,4$], to the Kerr
line element written in the same approximation.

 The generality of the mass formula $(4)$
and its crucial disagreement with $(15)$ allow us to make a generic statement:
stationary rotating equlibrium  stellar configurations cannot
be a source of the Kerr metric.

A similar result has been recently obtained by S. Sengupta
[$6$], who suggests an interpretation of the parameter $a$
associated with
the Kerr metric as the specific angular momentum corresponding to the
unphysical situation  of a hollow
rotating object.
\medskip

{\bf Acknowledgement}

It is a pleasure to thank  E. Gourgoulhon for initiating an important
discussion, as
well as S. M. Chitre,  E. V. Chubarian, P. Ghosh  and S. Sengupta for
helpful comments.

\end{document}